\documentclass[fleqn,usenatbib]{mnras}

\usepackage[T1]{fontenc}

\DeclareRobustCommand{\VAN}[3]{#2}
\let\VANthebibliography\thebibliography
\def\thebibliography{\DeclareRobustCommand{\VAN}[3]{##3}\VANthebibliography}

\usepackage{graphicx}
\usepackage{amsmath}
\usepackage{amssymb}
\usepackage{siunitx}
\DeclareSIUnit\angstrom{\text {Å}}
\usepackage{newtxtext,newtxmath}

\graphicspath{{../}}

\title[PEGS~IV]{Precision Ephemerides for Gravitational-wave Searches -- IV: Corrected and refined ephemeris for Scorpius X-1}

\author[Killestein et al.]{
	T. L. Killestein,$^{1}$\thanks{E-mail: \href{mailto:t.killestein@warwick.ac.uk}{t.killestein@warwick.ac.uk}}
	M. Mould,$^{2}$
	D. Steeghs,$^{1,6}$
	J. Casares,$^{3,4}$
	D.~K. Galloway,$^{5,6,7}$
	J.~T. Whelan,$^{8}$
	\\
	$^{1}$Department of Physics, University of Warwick, Gibbet Hill Road, Coventry CV4 7AL, UK\\
	$^{2}$School of Physics and Astronomy \& Institute for Gravitational Wave Astronomy, University of Birmingham, Birmingham B15 2TT, UK \\
	$^{3}$Instituto de Astrofísica de Canarias, E-38205 La Laguna, Tenerife, Spain \\
	$^{4}$Departamento de Astrofísica, Universidad de La Laguna, E-38206 La Laguna, Tenerife, Spain\\
	$^{5}$School of Physics and Astronomy, Monash University, Victoria 3800, Australia \\
	$^{6}$OzGRav-Monash, School of Physics and Astronomy, Monash University, Victoria 3800, Australia \\
	$^{7}$Institute for Globally Distributed Open Research and Education (IGDORE) \\
	$^{8}$School of Mathematical Sciences and Center for Computational Relativity and Gravitation, Rochester Institute of Technology, Rochester, NY 14623, USA\\
}

\date{Accepted XXX. Received YYY; in original form ZZZ}

\pubyear{2023}

\begin{document}
	\label{firstpage}
	\pagerange{\pageref{firstpage}--\pageref{lastpage}}
	\maketitle
	
	\begin{abstract}
		Low-mass X-ray binaries have long been theorised as potential sources of continuous gravitational-wave radiation, yet there is no observational evidence from recent LIGO/Virgo observing runs. Even for the theoretically `loudest' source, Sco~X-1, the upper limit on gravitational-wave strain has been pushed ever lower. Such searches require precise measurements of the source properties for sufficient sensitivity and computational feasibility. Collating over 20 years of high-quality spectroscopic observations of the system, we present a precise and comprehensive ephemeris for Sco~X-1 through radial velocity measurements, performing a full homogeneous reanalysis of all relevant datasets and correcting previous analyses. Our Bayesian approach accounts for observational systematics and maximises not only precision, but also the fidelity of uncertainty estimates --- crucial for informing principled continuous-wave searches.  Our extensive dataset and analysis also enables us to construct the highest signal-to-noise, highest resolution phase-averaged spectrum of a low-mass X-ray binary to date. Doppler tomography reveals intriguing transient structures present in the accretion disk and flow driven by modulation of the accretion rate, necessitating further characterisation of the system at high temporal and spectral resolution. 
		Our ephemeris corrects and supersedes previous ephemerides, and provides a factor three reduction in the number of templates in the search space, facilitating precision searches for continuous gravitational-wave emission from Sco~X-1 throughout the upcoming LIGO/Virgo/KAGRA O4 observing run and beyond.
	\end{abstract}
	

	\section{Introduction}
	\label{sec:intro}
	With the completion of their third observing run, the catalogue of gravitational-wave (GW) signals detected by the Advanced Laser Interferometer Gravitational Wave Observatory (LIGO; \citealt[]{LIGOScientificCollaboration2015}) and Advanced Virgo~\citep{Acernese2015} is growing rapidly \citep{2021arXiv211103606T}. These sources include the mergers of binary black holes (BHs)~\citep{2016PhRvL.116f1102A}, a binary neutron star (NS)~\citep{2017PhRvL.119p1101A,2017ApJ...848L..12A}, and NS--BH binaries~\citep{2021ApJ...915L...5A}. However, compact-object mergers are not the only sources expected to produced detectable GW emission. Unlike these transient signals, continuous GW (CW) sources present persistent quasi-monochromatic emission. This as-yet undetected type of gravitational radiation is emitted by rapidly rotating asymmetric NSs, whether found as isolated sources or within stellar binaries; see, e.g., \cite{2015PASA...32...34L,2019Univ....5..217S,Piccinni2022} for recent reviews.
	
	Numerous mechanisms have been suggested to induce the time-varying quadrupole required for GW emission from spinning NSs, whose angular momentum loss would limit the spin rate and account for the observation that NS spins are measured at $\lesssim700~\mathrm{Hz}$~\citep{2003HEAD....7.1738H,2006Sci...311.1901H,2017ApJ...850..106P}, below estimated breakup frequencies~\citep{2003Natur.424...42C}. Low-mass X-ray binaries (LMXBs) with NS primaries have received particular attention as target sources~\citep{Abbott2007, 2015PhRvD..91j2005W, Aasi2015, Abbott2017, Abbott2017, Abbott2019, 2020PhRvD.102b3006M, Zhang2021, Abbott2022ScoX1HMM}. In this scenario, a torque balance is achieved between the spin-up due to accretion from a stellar companion and spin-down due to GW emission~\citep{1978MNRAS.184..501P,1984ApJ...278..345W,Bildsten1998,1999ApJ...516..307A}. Such a rotational equilibrium leads to a characteristic strain that increases with increasing X-ray flux of the source (a proxy for the accretion rate; see \citealt[]{Bildsten1998}). The most promising candidates are thus those that are most bright in X-rays.
	
	The prototypical LMXB, Sco~X-1~\citep{Giacconi1962,1966ApJ...146..316S,1967ApJ...148L...1S}, is composed of an accreting NS primary and donor star, and has been intensively studied since its discovery as among the closest X-ray binaries known \citep{1975ApJ...195L..33G,1999ApJ...512L.121B,Fomalont2001}. It shows strong emission across the electromagnetic (EM) spectrum, from radio \citep{Fomalont1983} to gamma-rays \citep{Brazier1990}, powered by a near-Eddington accretion rate from the donor star. Intriguing null detections of very high energy (VHE, TeV) emission \citep{Aleksic2011} suggest that the high-energy emission mechanism in Sco~X-1 is markedly different to other systems. The donor star in the system remains enigmatic; the high accretion luminosity shrouds any stellar absorption features present in the near-infrared \citep{MataSanchez2015}, but these observations combined with dynamical constraints suggest a donor mass of $0.28 < M_{2} < 0.7 \, M_{\odot}$ (stellar type K4IV or later). As the strongest source of X-rays on the sky, Sco~X-1 has been used as a `lighthouse' to study intervening interstellar material \citep{Garcia2011}, magnetic fields \citep{Titarchuk2001}, and even the Martian atmosphere \citep{Rahmati2020MarsSco}. Very recently, polarised X-ray emission was detected by the \textit{PolarLight} mission at keV energies \citep{Long2022}, providing a strong constraint on the system geometry and suggesting that the X-ray emission arises primarily from a compact, optically thin corona near the disc transition layer.
	
	Despite being a cornerstone of our understanding of compact binary systems across the Universe, Sco~X-1 still remains enigmatic, showing great complexity and variability. These complexities are simply not revealed in more distant systems where lower signal-to-noise ratio (SNR) limits their visibility. Putting this aside, Sco~X-1 is the most luminous extra-solar X-ray source, which combined with its relative proximity ($2.3 \pm 0.1$ kpc, \citealt[]{gaiaedr3_astrometric}), implies it should be among the loudest CW sources detectable by current GW detectors \citep{2008MNRAS.389..839W} and has made it the target of a great number of search efforts.
	
	No concrete detection of CW emission has yet been made of what we a priori expect is the strongest GW source, which remains puzzling. However, improvements in detector sensitivity have led to correspondingly stronger upper limits on the GW strain from Sco~X-1, \citep{Abbott2007,Aasi2015,Abbott2017,Meadors2017, Abbott2017CrossCorr, Abbott2019, Zhang2021,Abbott2022ScoX1HMM}, with the most recent analyses reaching $\lesssim4\times10^{-26}$ (in the most sensitive frequency bands and assuming knowledge of the inclination angle; \citealt[]{Abbott2022CrossCorrelation}). This level of strain begins to push below the torque-balance limit \citep{Bildsten1998}, where the expected GW emission balances the spin-up torque from donor accretion.
	One complicating factor in the case of Sco~X-1 is the absence of a measured spin period, unlike many other accreting NS systems \citep[]{Abbott2022AccretingMSP}. Sco~X-1 has also been the target of directional, stochastic searches \citep{Abbott2017Stochastic, Abbott2019Stochastic, Abbott2021Stochastic}, leading to further sensitive (yet also null) results.
	
	As one of the primary uncertainties in any directed search, extensive X-ray timing observations have been performed in search of the spin period of the (assumed) NS primary, as revealed by X-ray pulsations or bursts (e.g. \citealt[]{Galloway2010}). The most recent placed an upper limit of 0.034\% (90\% confidence) on any putative X-ray variability using Rossi X-ray Timing Explorer (RXTE) data \citep{Galaudage2022}. It remains unclear whether Sco~X-1 is intrinsically or intermittently variable in the X-ray, and whether any pulsed X-ray emission is being scattered away by surrounding material.
	
	Another crucial component to facilitate CW searches are precise orbital constraints for the LMXB systems of interest \citep{2008MNRAS.389..839W}. In order to combine long sections of data coherently, the motion of both the Earth relative to the Solar System and the orbital motion of the NS in the binary must be accounted for. It is computationally infeasible to marginalise over the vast, high-dimensional parameter space this presents, and therefore, it is critical to place constraints on this orbital motion to reduce the search dimensionality and increase sensitivity \citep{2001PhRvD..63l2001D,2015PhRvD..92b3006M,2015PhRvD..91j2003L}. This presents a signficant observational challenge however; in a high accretion rate system like Sco~X-1, the Balmer features are blurred significantly, limiting precision and making them unsuitable for precision radial velocity (RV) measurements. The disc structure is also dynamic and chaotic, leading to time- \emph{and} phase-dependent changes in the emission line geometry.
	
	Fortuitously, the irradiated donor star provides an alternative observational probe via narrow emission lines generated by Bowen fluorescence, which provide a remarkably precise probe of orbital motion, as first demonstrated by \citet{2002ApJ...568..273S}. The NS irradiates the face of the donor star with a strong X-ray flux, ionising He~II which then de-excites, triggering a cascade of emission from C/N/O atomic orbitals \citep{Kastner1996}. The (comparatively) compact emission geometry generates narrow emission lines (with little intrinsic broadening), which precisely trace the heated face of the donor in contrast to disc emission, which has contributions from either side of the disc and thus experiences significant Doppler broadening and complex geometric distortions from specific regions, such as the hot-spot/bulge or the gas stream. This tracer of orbital motion can be used to estimate the orbital parameters of the NS with RV measurements. Beyond Sco~X-1, this technique has found broad applicability to LMXBs in general (e.g., \citealt[]{Casares2003,Cornelisse2007,Brauer2018}). To this end, Sco~X-1 has benefited from an extensive spectroscopic monitoring campaign since 1999 as a part of the Precision Ephemerides for Gravitational-wave Searches (PEGS) project. This has lead to incremental improvements in the ephemeris accuracy \citep{Galloway2014, Wang2018}, with coverage continuing up to the present day through an extensive all-weather campaign with the Very Large Telescope (VLT) Ultraviolet and Visual Echelle Spectrograph (UVES; \citealt[]{Dekker2000}) providing over 200 high-quality spectra.
	
	In support of ongoing CW searches, in this paper, we present the most up-to-date ephemeris for Sco~X-1 in the literature. We leverage 20 years of spectroscopic coverage with a homogeneous approach that accounts for potential systematic errors --- with a view to maximising not only the precision of the ephemeris, but also the quality of uncertainty estimates --- to correct and improve upon previous analyses. This is crucial when constraining the parameter space for CW searches; over-optimistic predictions lead to missing out valid areas of parameter space, whereas over-estimated errors greatly increase the computational burden of such searches. In Section~\ref{sec: data reduction}, we describe the observational data used. In Section~\ref{sec:binary_ephemeris}, we describe the method to infer the orbital properties of Sco~X-1 via RV measurements and present our updated --- and corrected --- ephemeris. In Section~\ref{sec: binary properties}, we further explore the uncertain variability in Sco~X-1. We present our conclusions in Section~\ref{sec: conclusions}

	\section{Data reduction}
	\label{sec: data reduction}
	
	\subsection{Spectroscopic observations}
\begin{figure}
		\centering
		\includegraphics[width=\columnwidth]{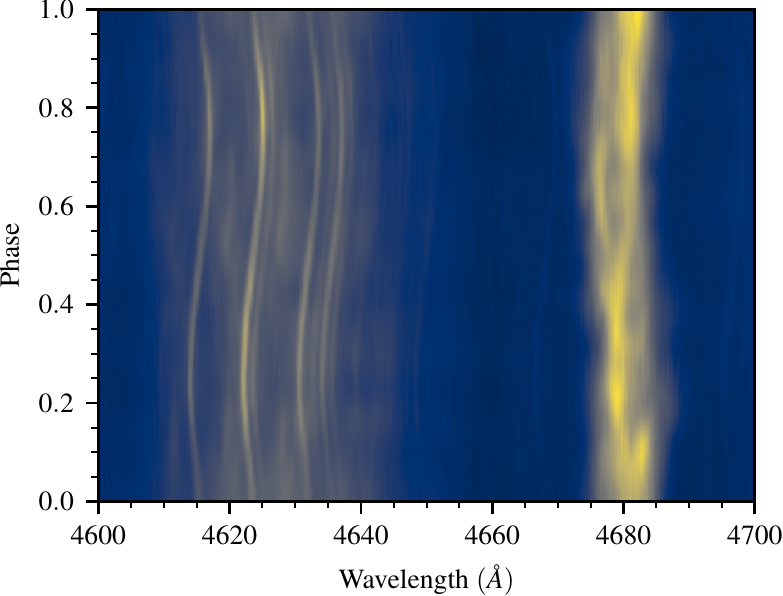}
		\caption{
			Trailed spectrogram of the VLT/UVES data presented in this work, folded by the best-fit ephemeris derived in Section~\ref{sec:binary_ephemeris}. 
			Centred on \SI{4640}{\angstrom} is the Bowen line region of interest, and \SI{4686}{\angstrom} is the He~II line.
			We restrict our analyses in this paper to this specific region of the spectrum, although other weaker Bowen lines are present in the spectrum.
		}
		\label{fig:trailedspectrogram}
	\end{figure}

	As part of the comprehensive analysis performed in this paper, we collate all spectroscopy presented in the previous PEGS project papers (\citealt[]{Galloway2014, Wang2018}) and combine this with more recent VLT/UVES data, extending the overall observational baseline to 22 years (1999-2021). We obtained (or retrieved) 264 spectra with the UVES \citep{Dekker2000} instrument mounted on the 8.2m VLT Unit Telescope 2. All observations were obtained in service mode, across variable observing conditions but with a typical SNR in the range 50--100. We focus here on the spectra taken with the blue arm using the \texttt{CD\#2} dispersive element, achieving a typical resolution $\Delta\lambda / \lambda\sim 40,000$ per spectral element with the 1" slit. Data were reduced using the v5.10.13 ESO UVES pipeline. We also include all relevant legacy datasets presented in the previous PEGS releases --- two runs of time-resolved spectroscopy taken with the ISIS double-beam spectrograph on the William Herschel Telescope (WHT) in 1999 and 2011, respectively. These observations were taken with the R400B grating and reduced using the \texttt{molly} software \citep{Marsh2019}. All spectroscopic datasets are enumerated in Table~\ref{tab: data}. A pictorial summary of the VLT/UVES data is given in Figure~\ref{fig:trailedspectrogram}, where we plot the phase-folded spectrogram of the Bowen line region implied by the binary constraints made in Section~\ref{sec: bayesian modelling}.
	
	\begin{table}
		\centering
		\begin{tabular}{c l l S[table-format=3.0]}
			\hline \hline
			Group & Instrument &     Program ID &  \multicolumn{1}{c}{No. spectra} \\
			\hline
			1 &   VLT/UVES &  077.D-0384(A) &                  5 \\
			1 &   VLT/UVES &  087.D-0278(A) &                 52 \\
			1 &   VLT/UVES &  089.D-0272(A) &                 57 \\
			1 &   VLT/UVES &  098.D-0688(A) &                 32 \\
			1 &   VLT/UVES &  599.D-0353(A) &                 29 \\
			1 &   VLT/UVES &  599.D-0353(B) &                 89 \\
			2 &   WHT/ISIS &  W/1999A/CAT42 &                137 \\
			3 &   WHT/ISIS &    W/2011A/P23 &                157 \\
			\hline
			&            &                &                558 \\
			\hline
		\end{tabular}
		\caption{Summary table containing all observations included in this ephemeris version, along with instrument and program IDs. `Group' here refers to the fitting group introduced in Section~\ref{sec: bayesian modelling} --- datasets in the same group are fitted with common error scaling and velocity offset parameters.
		}
		\label{tab: data}
	\end{table}
	
	For all datasets, we recompute both the barycentric time and velocity corrections at mid-exposure to reduce any scatter introduced by the different conversion routines used in the data processing pipelines. We use the \texttt{astropy.time} module and adopt the JPL DE405 ephemerides as our reference. To provide both consistency with previous ephemerides and an authoritative value, we give ephemeris times in both UTC Barycentric Julian Date (BJD) and GPS seconds.

	\subsection{Measuring Bowen line velocities} \label{sec:linemodel}
	\begin{figure}
		\centering
		\includegraphics[width=\columnwidth]{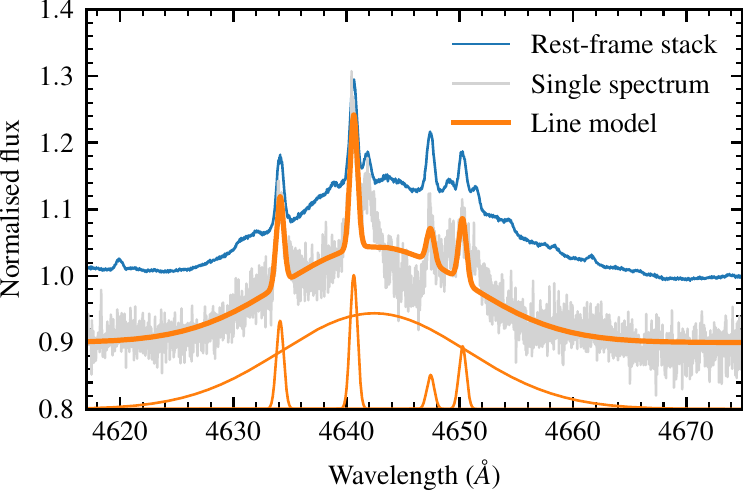}
		\caption{ Example plot of the Bowen region in one UVES spectrum after continuum subtraction, with the Bowen line model discussed in Section~\ref{sec:linemodel} overplotted along with the individual components. Note that one of the fitted line components is not visible in this spectrum so has zero amplitude.
			We take care to avoid including the bright HeII~\SI{4686}{\angstrom} line. Individual spectra typically have SNR $\sim$ 50.
		}
		\label{fig:rawspectra}
	\end{figure}
	
	We obtain our Bowen radial velocities by fitting a constrained line model similar to the one used in \citet{Wang2018} to the reduced spectra. An example spectrum with the model is given in Figure~\ref{fig:rawspectra}. The narrow Bowen components are modelled with Gaussians of fixed width (\SI{50}{\kilo\metre/\second}) and variable amplitudes, offset from their rest frame with a common velocity. In the region of interest, we identify five narrow-line N~III/C~III/O~II components that optimally constrain the Bowen RVs:
	\SI{4634.13}{\angstrom},
	\SI{4640.64}{\angstrom},
	\SI{4647.42}{\angstrom},
	\SI{4650.25}{\angstrom}, and
	\SI{4643.37}{\angstrom} respectively.
	Some of these narrow lines are not visible at specific phases due to system geometry. To accommodate this, we constrain all line amplitudes to be $\geq$ 0, such that these emission lines cannot be forced by continuum modulation to unphysical non-negative fluxes when they are not present. The broad-line component is modelled with a Gaussian of fixed width \SI{1250}{\kilo\metre/\second}. The amplitude and centroid of this component are fitted for to remove additional correlations with the centres of the narrow lines. The line widths above were measured by Gaussian fits to the individual components, and are kept fixed to limit the number of free parameters. Empirically, the narrow-line components do not change width significantly as a function of orbital phase, and any model--data mismatch is unlikely to cause large deviations in the fitted velocities as the emission-line peaks are typically clear and share a common velocity.
	
	The final line model has eight free parameters that are well constrained by the dense spectral sampling ($\approx$ \SI{0.03}{\angstrom}/pix) of the UVES data. Our specific line model mitigates correlations between parameters, such that uncertainties in the other fitted parameters do not skew the Bowen line velocities. Through experimentation, the UVES data could potentially benefit from the addition of more lines, but this leads to issues in the lower-quality WHT data with deviant line amplitudes so we opt for the above model for all datasets.
	
	As a pre-processing step, we continuum-subtract the spectra by fitting a third-order Chebyshev polynomial to \SI{10}{\angstrom} regions at the red and blue ends of the Bowen line region (\SIrange{4605}{4675}{\angstrom}). We fit the line model using nonlinear least squares, as implemented in the \texttt{least\_squares} routine in the \texttt{scipy} package \citep{scipy}. The original ephemeris of \citet{Wang2018} is used to set the initial Bowen line velocity, and other parameters are initialised with sensible defaults from the highest SNR spectrum. The RVs are then computed from the mutual Doppler shift of the measured emission-line locations with respect to their known rest-frame values. Uncertainties are rescaled to enforce a unity reduced chi-squared test statistic. We add a constant value of \SI{0.5}{\kilo \metre / \second} in quadrature with the error values to ensure uncertainties are not underestimated due to poor conditioning of the least-squares fit, and to include uncertainties in the absolute velocity calibration of UVES. At this stage of reanalysis we are more concerned with correct \textit{relative} error scaling between velocity measurements, as we rescale the errors between each dataset in latter analysis steps.
	
	\section{Binary ephemeris}
	\label{sec:binary_ephemeris}
	
	\subsection{Corrections to previous Sco~X-1 ephemerides}
	\label{sec:corrections}
	
	As part of our homogeneous reanalysis, we identified calibration errors that affect the timing parameters of both previously published ephemerides for the Sco~X-1 system \citep{Galloway2014, Wang2018}. These were not readily apparent in previous work as the statistical uncertainties on the WHT-derived datasets masked their effect on the ephemeris. With the inclusion of four times as many VLT spectra, these discrepancies become significant and thus must be corrected. For full transparency, we elaborate in greater detail on the specific errors and their effect on the ephemeris:
	\begin{itemize}
		\item The subset of VLT data used in PEGS~I and PEGS~III were inadvertently not corrected to the mid-exposure time, nor adjusted to the Solar System heliocentre. As a result, the values of $T_0$ presented in these works bear systematic errors of $\sim$ 600s.
		\item Covariances between orbital parameters are underestimated in \citet{Wang2018} due to being taken from the initial least-squares fit to the data (using a two-point finite-difference Jacobian approximation), rather than being computed from the samples from which the ephemeris is derived. This changes the covariance by two orders of magnitude.
	\end{itemize}
	
	Discrepancies were identified by cross-checking the reduced data against the raw frames located at the European Southern Observatory (ESO) Archive, in particular the metadata stored in the file headers. Through a careful reanalysis of the datasets, we confirm that addressing these issues brings the data used in both G14 and W18 into strong agreement, and jointly agree with the VLT data presented in this paper (accounting for deviations in the period due to dependence on low-resolution WHT data). We plot marginal posteriors for each ephemeris in Figure~\ref{fig:ephemeris_fixes} to illustrate how the ephemerides change with the corrections applied above. These issues underscore the importance of both robust archiving of astronomical data, such that raw frames are easily retrievable even decades later, and the importance of storing metadata alongside these frames and documenting reduction steps; without this, these timing issues could not have been easily diagnosed. With this additional scrutiny, we are now confident any calibration artifacts are accounted for in our reanalysis -- our previously-published ephemerides are considered obsoleted by the ephemeris presented in this paper, and should not be used in CW searches going forward.
	
	\subsection{Bayesian modelling of the Keplerian orbit}
	\label{sec: bayesian modelling}
	\begin{table}
		\centering
		\begin{tabular}{c l l}
			Parameter & Value & Units\\
			\hline \hline
			$T_0$ & $2456723.3272 \pm 0.0004$ & (BJD UTC)\\
			& $1078170682 \pm 33$ & (GPS seconds) \\
			$T_{\mathrm{asc, ns}}$ & $1078153676 \pm 33$ & (GPS seconds) \\
			$P$ & $0.78731387 \pm 0.00000019$ & (days) \\
			& $68023.919 \pm 0.017$ & (seconds) \\
			$K$ & $76.8 \pm 0.2$ & (km/s) \\
			$\gamma$ & $-113.8 \pm 0.2$ & (km/s) \\
			\hline
			$e$ & $\leq 0.0132$ & \\
			\hline
		\end{tabular}
		\caption{Tabulated posterior distribution summary statistics for our ephemeris. Values correspond to the posterior median, while uncertainties represent the marginal $1\sigma$ confidence intervals. Explicitly, the value of $T_0$ refers to the inferior conjunction of the companion star, and $T_{\mathrm{asc, ns}}$ refers to the time of ascending node crossing for the NS, moving away from the observer.
			The top rows assume zero eccentricity. Our upper limit on the eccentricity is given in the bottom row, as the 90th percentile of the marginal eccentricity distribution. Accompanying this table is a corner plot of posterior samples in Figure \ref{fig:jointcornerplot}.}
		\label{tab:the_ephemeris}
	\end{table}
	
	\begin{figure}
		\centering
		\includegraphics[width=\linewidth]{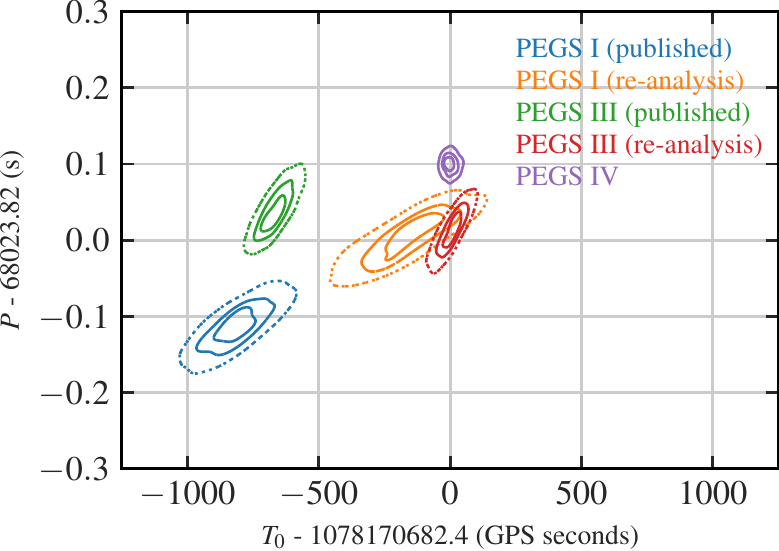}
		\caption{Marginal two-dimensional posteriors of $T_0$ and $P$ for all previous ephemerides, propagated forwards to the epoch of our ephemeris (see Table\ref{tab:the_ephemeris}). The contours show the 1-, 2-, and 3-$\sigma$ confidence intervals respectively.}
		\label{fig:ephemeris_fixes}
	\end{figure}
	
	\begin{figure*}
		\centering
		\includegraphics[width=\linewidth]{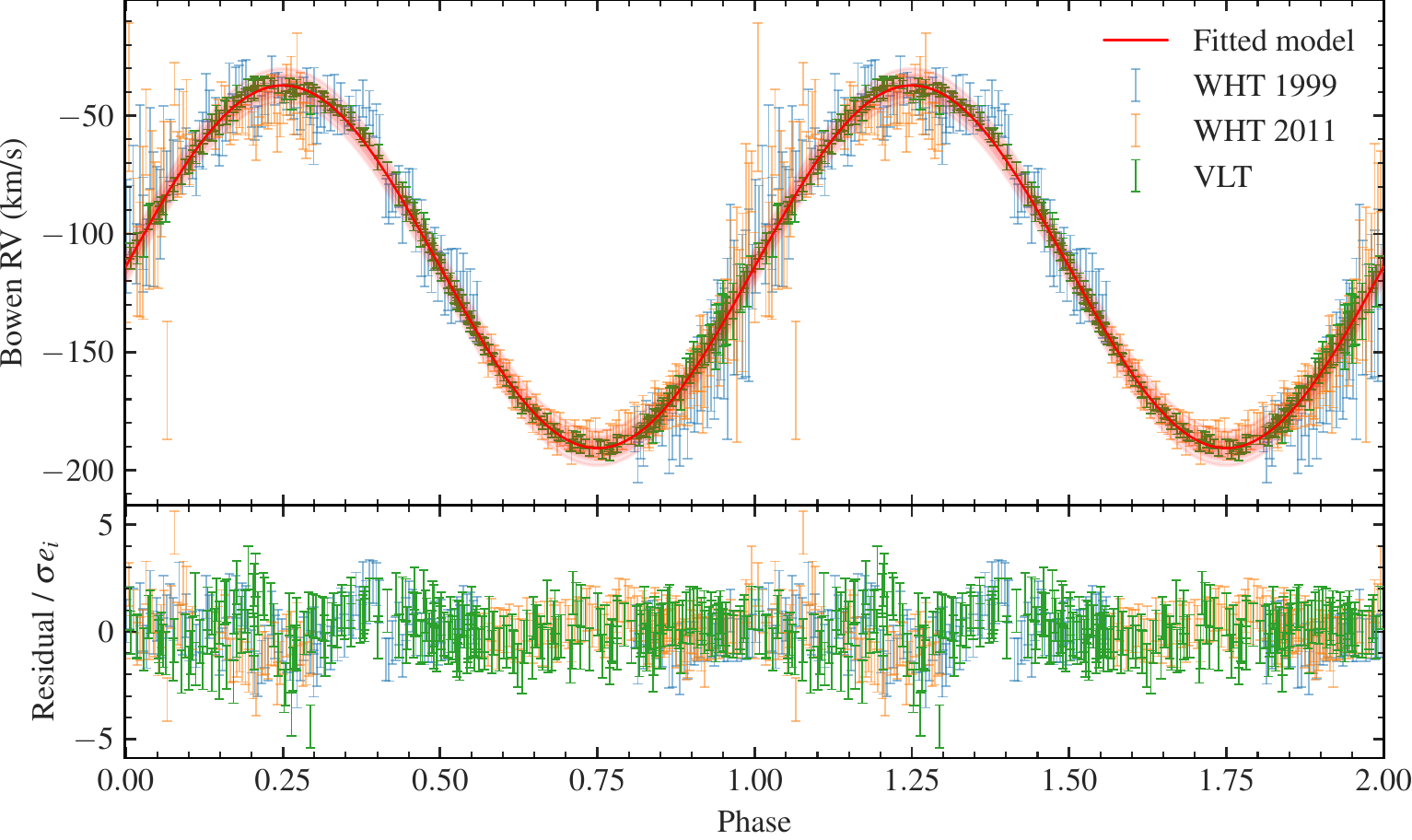}
		\caption{Top panel: RV curve for Sco~X-1. We show the $1\sigma$ confidence intervals of the RVs measured from the 1999 WHT (blue), 2011 WHT (orange), and VLT (green) spectroscopic datasets, derived from the Bowen line model. We also plot the median posterior predictive RV curve implied by our ephemeris constraints (red) and the corresponding $1\sigma$, $2\sigma$, and $3\sigma$ confidence regions (darker to lighter red shaded regions, respectively). Note the significantly reduced intrinsic scatter of the higher resolution VLT data with respect to the older WHT data.
			Bottom panel: Normalized residuals between the observational RV measurements with the median RV curve, scaled by the respective error factors $e_i$}
		\label{fig:joint_rv_model_fit}
	\end{figure*}
	
	Moving forward with the reanalysis, we apply here the velocity corrections derived during the previous steps to bring all spectra into the Solar System barycentric frame, and convert observation times to BJD.
	
	We derive our ephemeris using a standard Keplerian orbit model (assuming zero eccentricity), using for the Bowen line RVs
	$$
	v(t) = K \sin\left(\frac{2\pi(t - T_0)}{P}\right )+ \gamma
	\, ,
	\label{eqn:rvmodel}
	$$
	where $K$ is the velocity semi-amplitude, $T_0$ is the reference epoch, $P$ is the period, and $\gamma$ is the systemic velocity. To ensure good initialisation prior to sampling, we first do a simple fit to the above model with nonlinear least-squares. To minimise covariance between $P$ and $T_0$ in our ephemeris, we use a simple bracketing line search to find the time of conjunction $T_0'=T_0+nP$ that minimises $\mathrm{cov}(P,T_0)$ for an integer number of orbital cycles $n$. This `seed' ephemeris is then used to initialise parameters for the more complex modelling that follows, reducing burn-in time during sampling and ensuring our final ephemeris provides the minimal covariance.
	
	To marginalise over systematics, we include two nuisance parameters per dataset: a constant offset term, $\delta_i$, to correct for differences in absolute wavelength calibration between the datasets, and an error scaling term, $\epsilon_i$, intended to correct for underestimated uncertainties. We assume a Gaussian likelihood on the RVs $\mu$ and their uncertainties $\sigma$ measured at time $t$, given by
	$$
	\log\mathcal{L}(\mu,\sigma,t|T_0,P,K,\gamma,\delta_i,\epsilon_i)
	\propto
	\sum_i \left( \frac{\mu-v(t)-\delta_i}{\epsilon_i\sigma} \right )^2
	\, ,
	$$
	where $i$ represents the index of each individual dataset, $\delta_i$ represents the per-dataset velocity offset, $\epsilon_i$ represents the per-dataset error scaling. The VLT datasets have the most stable and accurate long-term calibration, and so the $\delta_i$ value for this dataset is fixed to zero. It should be noted that this offset term effectively also absorbs long-term velocity variations in the system --- e.g., due to perturbations from a distant companion. Over the observational baseline, these would be linear in order. Regardless, the focus of this work is to provide the most precise timing of the system possible, and we have no reason to expect significant secular motion. To avoid being biased by previous ephemerides calculated from these datasets, we assume uninformative uniform priors on all variables. To mitigate complications associated with multimodality arising from an unconstrained $T_0$, the prior range of $T_0$ is centred on the seed value and bounded on either side within the seed period.
	
	We generate samples from the posterior distributions with Hamiltonian Monte Carlo (HMC) sampling \citep{Duane1987, Betancourt2017} and the No U-Turn Sampler (NUTS; \citealt[]{Hoffman2011}), using \texttt{JAX} \citep{jax2018github} and \texttt{NumPyro} \citep{phan2019composable, bingham2019pyro}. The number of burn-in and sampling steps are tuned to ensure convergence via the Gelman-Rubin split-$\hat{r}$ convergence diagnostic \citep{Gelman1992}. We sample four chains in parallel for 4000 steps in total, discarding the first 2000 samples of each chain as burn-in. This procedure takes just two minutes to complete on commodity hardware, running a single chain per core.
	
	As a final step to further minimise $\mathrm{cov}(P,T_0)$, we repeat the process of shifting $T_0$ by an integer number of periods, this time using the HMC samples to provide a more robust estimate of the covariances. We find a shift of -106 periods minimises this, and thus we adopt this value as $T_0$ going forward.
	
	Our updated ephemeris is presented in Table~\ref{tab:the_ephemeris}. Compared to (our corrected version of) the ephemeris from PEGS~III, we achieve an factor 2.9 improvement in the uncertainty on the time of conjunction when propagated to the start of O4. In the top panel of Figure~\ref{fig:ephemeris_fixes}, we compare the two-dimensional marginal distribution of $(T_0,P)$ for our updated ephemeris with that for each of the previous (corrected and uncorrected) ephemerides, propagated to the epoch of the new ephemeris. Our corrections are clear in the now consistent values of the reference epoch, $T_0$. There are some deviations in the period $P$, however this is expected due to the inclusion of more VLT data (with better resolution of the spectral lines) at later epochs.
	
\begin{figure*}
		\centering
		\includegraphics[width=\linewidth]{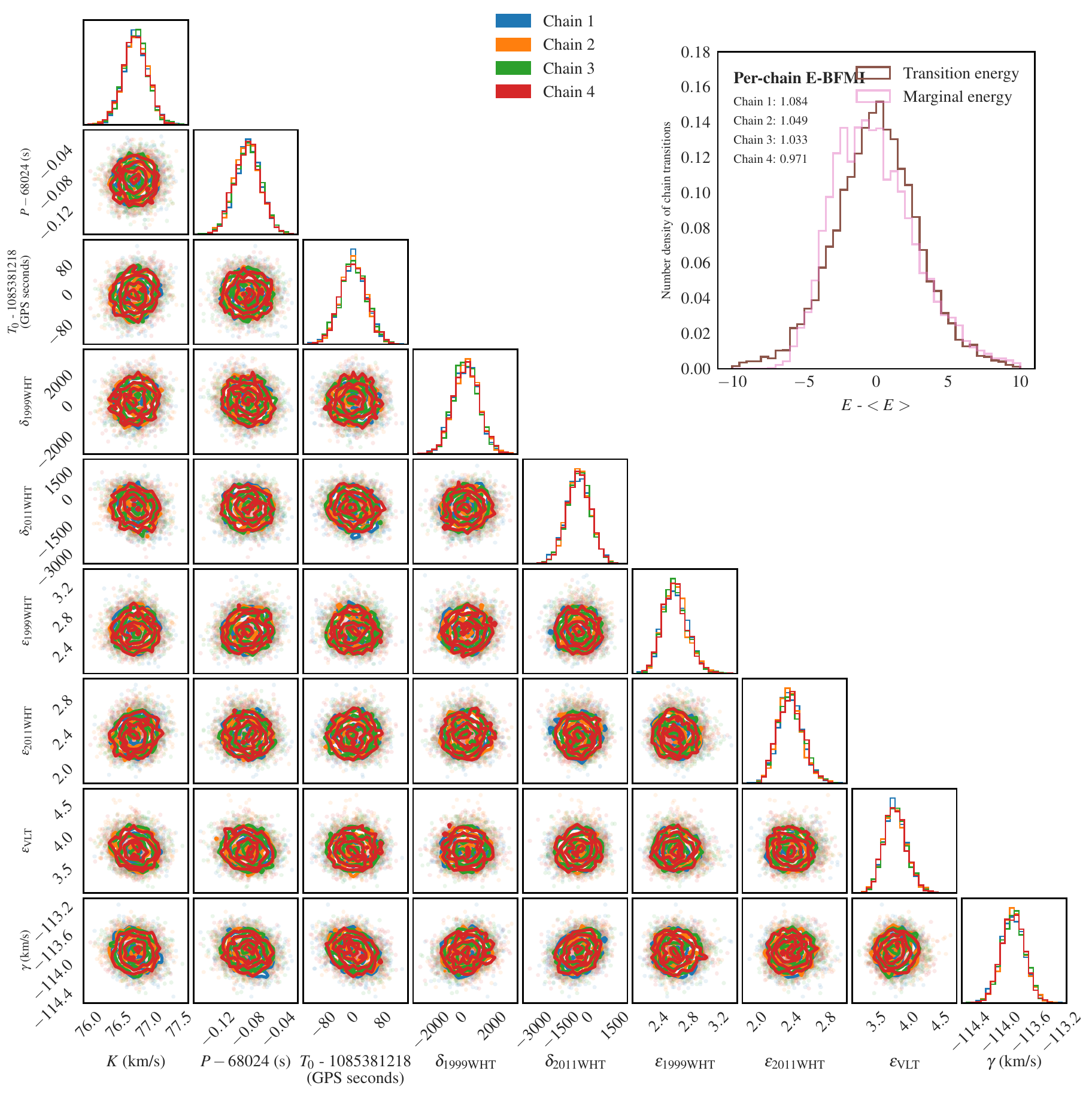}
		\caption{Corner plot of posterior samples for the $e=0$ ephemeris, coloured according to each independent chain to confirm randomly-initialised chains converge on the same posterior mode. Inset: Histogram showing the distributions of marginal energies against the transition energies for each step taken by the NUTS sampler - the close match between these two distributions implies full and efficient (low autocorrelation) exploration of the parameter space. The estimated Bayesian fraction of missing information is close to 1 for all chains, providing a quantitative verification of this also. These diagnostics are unique to Hamiltonian Monte Carlo and provide an orthogonal check to the usual split-$\hat{r}$ diagnostics employed in MCMC that we use above. Refer to \citet{Betancourt2016} for a full theoretical explanation, and further details.}
		\label{fig:jointcornerplot}
	\end{figure*}

	In the bottom panel of Figure~\ref{fig:ephemeris_fixes}, we propagate the uncertainty on the PEGS~IV $T_0$ to past and future epochs. This again demonstrates the consistency of our corrected and new ephemerides with each other. Of course, the uncertainty in the reference epoch $T_0$ grows in time, due to the posterior uncertainty in the ephemeris. However, our updated measurements reduce this rate of growth into the future observing runs, O4 and O5, of the LIGO/Virgo/KAGRA GW detectors, as is crucial for increasing the sensitivity of searches for CWs from Sco~X-1.
	In Figure~\ref{fig:joint_rv_model_fit}, we present the RV curve of Sco~X-1 implied by our ephemeris inference. The high-resolution VLT spectra result in an excellent match with the fitted RV curve, with robust uncertainty quantification carried through our Bayesian measurements of the ephemeris. Finally, Figure~\ref{fig:jointcornerplot} presents a corner plot of our samples, along with diagnostic statistics to illustrate the proper exploration of the parameter space.
	
	For completeness, we present the ephemeris posterior and additional diagnostic quantities in Figure~\ref{fig:jointcornerplot}. To enable the principled calculation of search boundaries based on the full posterior distributions, we make high-quality samples available alongside this paper in the Supplementary Material section.
	
	\begin{figure}
		\centering
		\includegraphics[width=\linewidth]{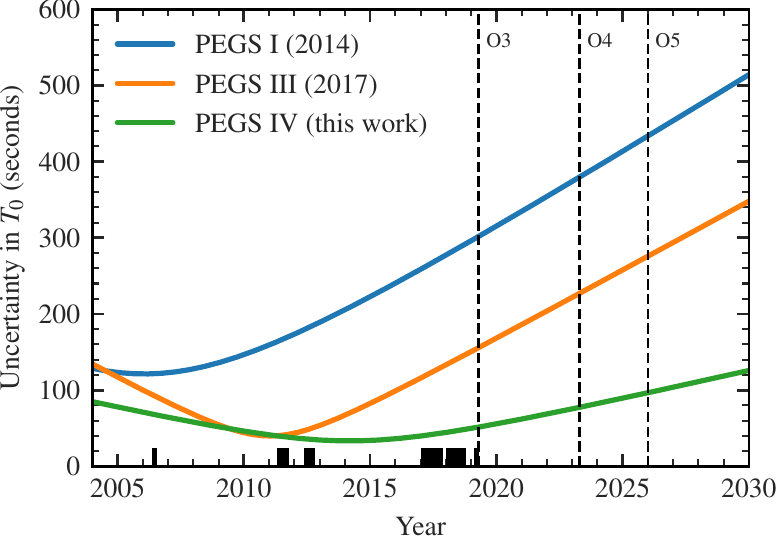}
		\caption{Plot of uncertainty in $\mathrm{T_0}$ as a function of epoch for all updated ephemerides presented in this work. The black bars overplotted denote the times of individual spectra contributing to the ephemeris. Our ephemeris provides a factor 2 improvement in uncertainty for the O3 observing run, growing out to later times. }
		\label{fig:ephemeris_propagation}
	\end{figure}
	
	Examining the posteriors in Figure~\ref{fig:jointcornerplot}, our marginal distributions are largely Gaussian, and inter-parameter correlations are very weak. 
	Our two-step line-search algorithm has successfully reduced the covariance between $T_0$ and $P$ significantly to an (absolute) value of $3.6 \times 10^{-15}\,\mathrm{d^2}$ -- multiple orders of magnitude improvement over the value quoted in \citet{Galloway2014}.
	It is further reassuring to see that both $\delta_{1999 WHT}$ and $\delta_{2011 WHT}$, the systemic offset parameters for each dataset, are consistent with zero at the 1$\sigma$ confidence level, verifying the validity of the absolute wavelength calibration of both datasets -- nevertheless their inclusion is important to account for a potentially non-trivial systematic effect on the ephemeris, and to provide correct uncertainty estimates on each parameter.
	
	Our ephemeris provides an excellent ($\sim 30$s) precision on the time of conjunction $\mathrm{T_0}$, providing a significant ($\sim$ factor 3) reduction in the required template search space (following \citealt[]{Galloway2014}). At these levels of precision, even subtle effects begin to have marked impacts on the ephemeris. It is  therefore crucial to acknowledge that there are additional sources of statistical and systematic uncertainty on timings that may hamper efforts to push the ephemeris to even greater precision.
	
	Sco~X-1 shows optical variability on the level of $\sim$ 0.5 mag, with typical variability timescales of $\sim 1$ hour. As a result, we observe a flux-weighted average radial velocity over the length of our exposure, adding additional uncertainties to our velocity estimates. This is likely a small effect at the sub-\SI{}{\kilo\meter/\second} level owing to our comparatively short exposure times ($\lesssim 5\%$ flux variability over one $\sim$ 700s UVES integration), however may begin to be important when searching for deviations from Keplerian radial velocities --- we discuss this further in Section~\ref{sec:eccentricity}.
	There are also potential uncertainties arising from the timestamps applied to each spectrum acquired -- it is challenging to quantify the temporal accuracy as this is often poorly documented, and rarely validated experimentally.  This is particularly true in the case of `historic' datasets from many decades ago. In the ideal scenario, timestamps would be derived from GPS time (e.g. see \citealt[]{Dhillon2007}), or derived from system time that is updated regularly via Network Time Protocol (NTP), each of which can keep clock errors minimal. This also implies potential per-observatory time offsets, which may manifest in non-trivial ways. These effects are currently dwarfed by the statistical error present on our ephemeris, and are only likely to become problematic given a significantly increased volume of data -- which will reduce the statistical errors on the ephemeris.
	We nevertheless caution that the `true' uncertainties on our ephemeris may be larger than implied by our posterior samples, and encourage conservative allocation of CW parameter spaces to accommodate this. Future work will look to incorporate some of these effects into our Bayesian framework, as well as further extending the observational baseline to provide the volume of data required to model them.
	
	\subsection{Eccentricity constraints} \label{sec:eccentricity}
	
	We now consider potential deviations from the standard circular orbit model discussed above. As remarked upon in \cite{Wang2018}, obtaining direct constraints on the orbital eccentricity of Sco~X-1 is complicated by the Roche lobe geometry of the companion star. As it fills its Roche lobe, emission occurs from an extended surface of the donor, which continuously changes aspect ratio with orbital phase. This naturally leads to deviations from the pure sinusoidal RV curve expected from the base Keplerian orbit model, and adds additional periodic modulation on the $\sim$\SI{1}{\kilo\metre/\second} level. This creates an apparent eccentricity dominating over --- and potentially degenerate with --- any true orbital eccentricity, which makes it difficult to disentangle the effects of each. As with other LMXBs, there are strong physical reasons to expect the orbital eccentricity $e$ of the system to be close to $e=0$, due to tidal dissipation occurring during the main sequence lifetime of the NS progenitor (see, e.g., \citealt[]{Tassoul1992}). As a further complication, the vanishing phase space \citep{Lucy1971} as $e\rightarrow0$ induces a bias towards non-zero orbital eccentricities, and therefore, any detection of $e>0$ should be interpreted with caution mathematically also. 
	
	Bearing this in mind, we repeat the analysis of Section~\ref{sec: bayesian modelling}, but fit for a nonzero eccentricity, using a more general form of the Keplerian orbital model described previously. We adopt the common parameterisation $(\sqrt{e}\cos{\omega}, \sqrt{e}\sin{\omega})$ to sample the orbital eccentricty $e$ and the argument of periapsis $\omega$. This decorrelates the posterior at low eccentricities and make sampling easier, assuming uniform priors on both of these parameters in the range (-1, 1). We use the \texttt{JAX}-based \texttt{kepler} solver from the \texttt{exoplanet} package \citep{Foreman-Mackey2021} for compatibility with the NUTS algorithm.
	
	We empirically find an upper limit on the orbital eccentricity of Sco~X-1 of 0.0132 (0.0161) at 90\% (99\%) confidence level.
	The commonly-employed significance test of \cite{Lucy1971} is of limited utility here as it considers solely orbital eccentricity, and (falsely) suggests we should adopt an elliptical orbit as a result. We expect any non-zero eccentricity originates entirely from the Roche geometry, bearing in mind the tidal dissipation seen in similar LMXBs. This provides a conservative constraint of parameter space for potential CW searches; it is difficult to move beyond this without more intensive modelling efforts, involving full modelling of the binary system in both light curves and RVs. Some early progress is being made (see, e.g., \citealt[]{Cherepashchuk2021}), although it is important to note that light curve modelling carries potential systematics that must be carefully accounted for when combined with other independent constraints. Even with the high-quality UVES data here, the expected RV modulation from the distorted secondary (see Figure~6 of \citealt[]{Wang2018}, of order \SI{100}{\metre/\second}) is comparable to the statistical uncertainty, making the prospect of direct measurement unlikely --- the greatest modulation occurs around phase zero, where the Bowen emission lines are weakest. Deviations from a pure sinusoid are directly informative on the inclination $i$ \citep{Masuda2021}, making further reduction of RV uncertainties an important goal going forward.

	\section{Binary properties}
	\label{sec: binary properties}
	As the prototypical LMXB, Sco~X-1 has been the focus of intense study since its discovery in the early 1960s \citep{Giacconi1962, Chodil1965}. However, the high (super-Eddington) accretion rate and its effect on the Balmer lines means there is still uncertainty surrounding the structure of the accretion disc in Sco~X-1 --- whether the observed form is stable over long timescales, and how this correlates with the various X-ray and optical states of the system (as identified in \citealt[]{Scaringi2015}). The long-term, high-resolution dataset collected as part of the PEGS program provides the means to probe secular evolution of the disc, as well as search for period derivatives and other phenomena not detectable with the typical dense but short-baseline observational sampling presented in previous work, although one pays a penalty in the resolution of fine structure owing to the dynamic and changing disc state. We expand on this in the subsections below.
	
	\subsection{High-resolution, high-SNR line atlas for Sco~X-1}
	\label{sec:lineatlas}
	With over 200 high-resolution spectra of Sco~X-1, and a high-precision ephemeris, we construct a `line atlas' by combining the spectra in the donor-star rest frame. By doing this, weak spectral lines not visible in the individual spectra can be detected, and the quality of all lines improved. The variability in the broad H/He line profiles leads to some smearing of these lines. However weaker, narrow lines are not affected as heavily. A significant number of VLT/UVES spectra taken as part of PEGS also have a simultaneous observations with the red arm, which, although of limited utility for constraining the binary motion due to the lack of strong and narrow spectral features, encodes information about the cooler material present in the Sco~X-1~ system. We include this in our line atlas primarily for completeness, although caution that telluric subtraction was not performed as part of data reduction, and so the red arm suffers from residual telluric contamination. This is somewhat suppressed by our shift-adding scheme.
	
	We use the measured Bowen line velocities to shift all spectra into alignment, and normalise out the continuum flux by fitting Chebyshev polynomials (7th order for the UVES Blue arm, 5th for UVES Red arm CCD 1, 1st for UVES Red arm CCD 2), rejecting outlier points to avoid fitting spectral features. The heavy telluric contamination redwards of \SI{9000}{\angstrom} necessitates the use of a low-order polynomial to avoid erratic ringing effects. Spectra are then reinterpolated onto a common wavelength grid, and median-combined with a sigma-clipping outlier rejection scheme to remove any cosmic ray hits or defects. This process is performed separately for the blue arm and the red arm, and we take care to treat the two red CCDs separately to avoid discontinuities. Note that the continuum normalisation at the far-red end of the spectrum is hampered by the strong telluric bands, so we opt for a simple linear continuum only. 
	
\begin{figure*}
		\centering
		\includegraphics[width=\linewidth]{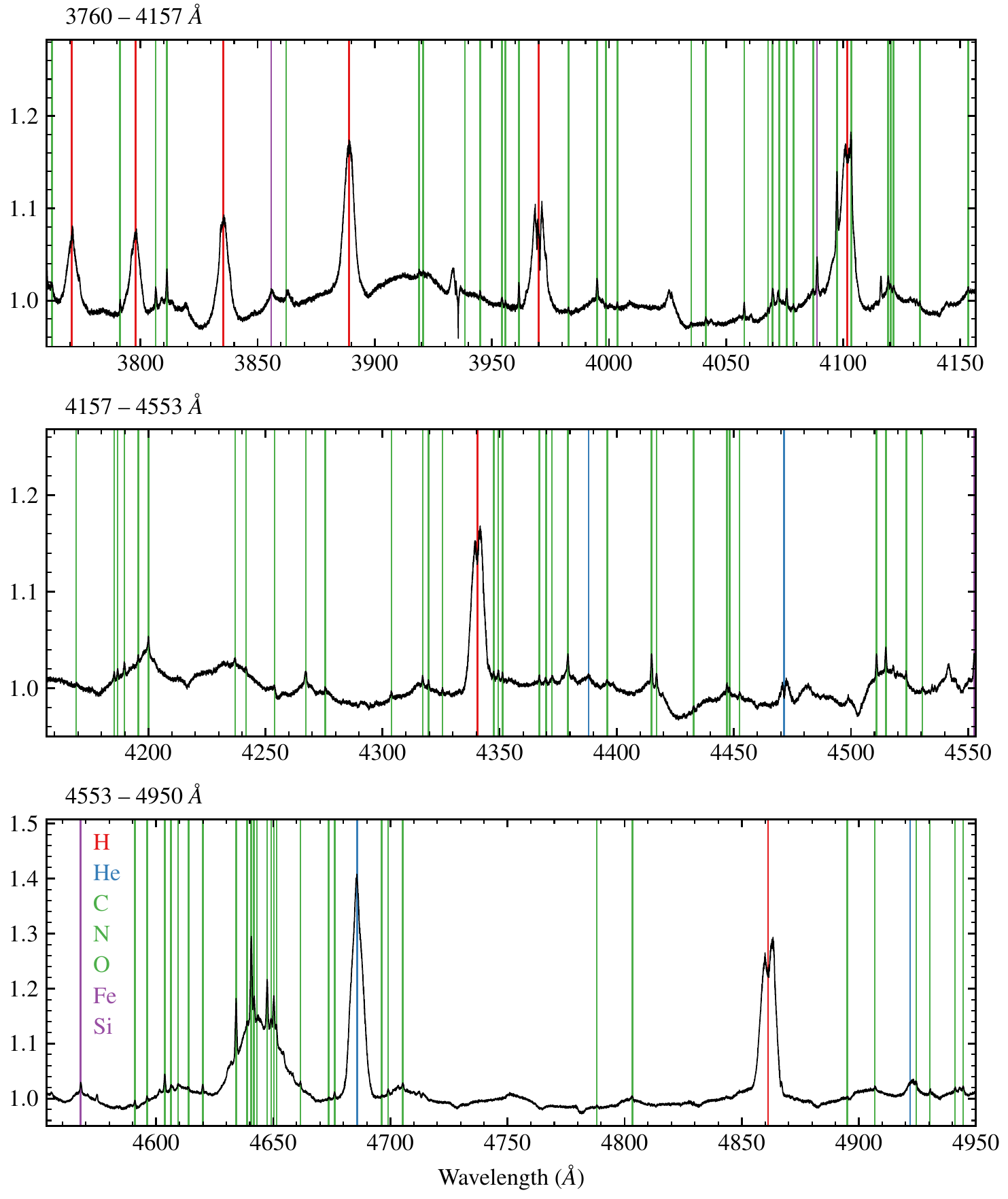}
		\caption{Co-addition of all VLT/UVES spectra in the binary rest-frame, to produce a line atlas that reveals the presence of many low-strength emission lines. Spectra are broken into chunks of \SI{400}{\angstrom} to aid visualisation. We annotate potential lines of interest to the LMXB community.}
		\label{fig:uves_spectral_atlas_blue}
	\end{figure*}

\begin{figure*}
		\centering
		\includegraphics[width=\linewidth]{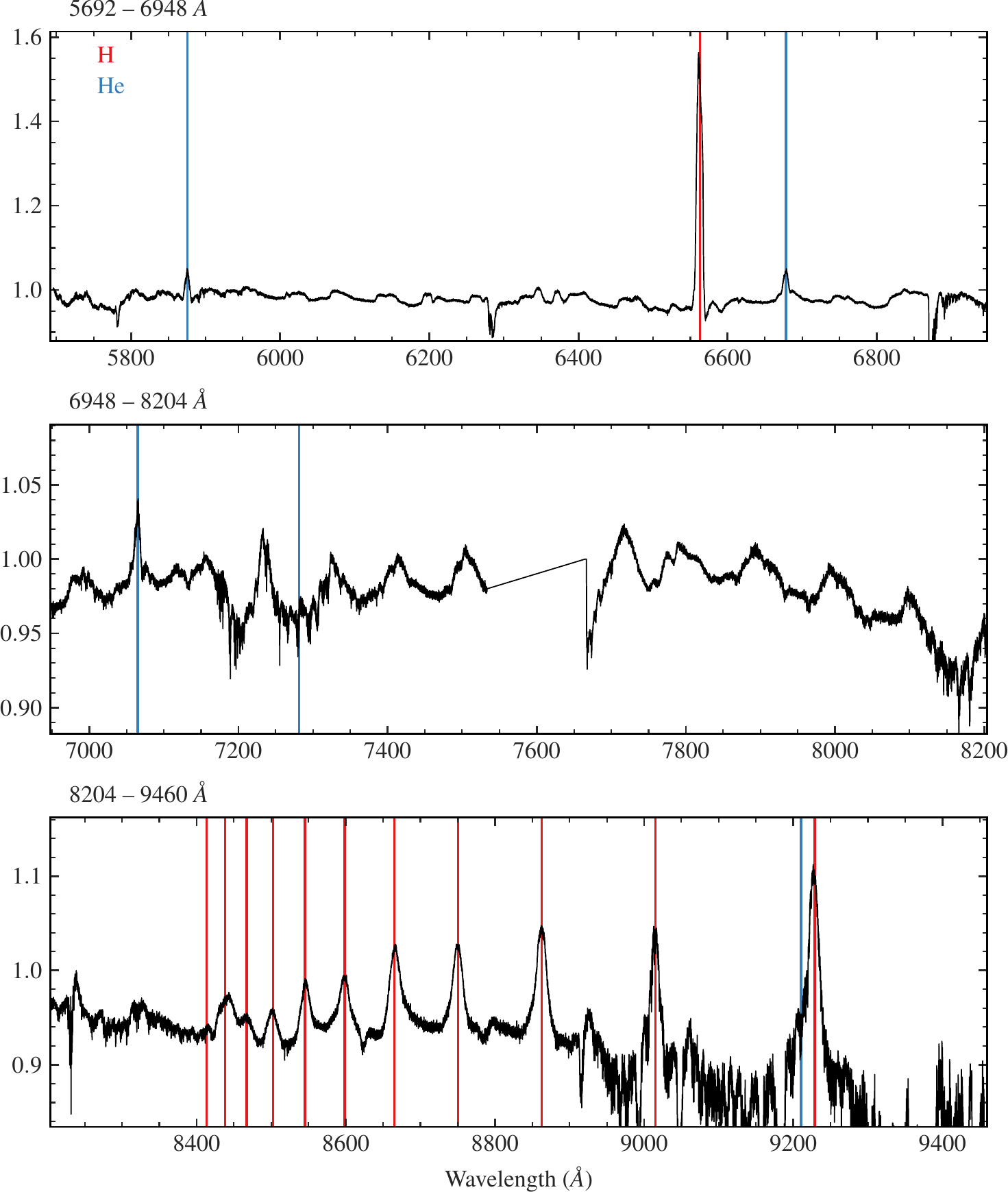}
		\caption{Line atlas for Sco~X-1~in the binary rest frame continued from Figure \ref{fig:uves_spectral_atlas_blue}~--~this figure shows the UVES red arm spectra, covering \SIrange{5692}{9460}{\angstrom}. Note that no telluric correction has been applied, and we instead use the binary motion to median the weaker features out. Some contamination persists in the red, hence we only attempt to identify the strong spectral lines of H and He, with the weaker C/N/O lines obscured by atmospheric absorption. A full treatment of telluric absorption is required to identify all faint lines present, but this is beyond the scope of this paper.
			UVES has a chip gap covering approximately \SIrange{7540}{7660}{\angstrom} in this configuration. Spectra are broken into chunks of \SI{1200}{\angstrom} to aid visualisation.}
		\label{fig:uves_spectral_atlas_red}
	\end{figure*}

	Our finalised line atlas is presented across Figure \ref{fig:uves_spectral_atlas_blue} (CD\#2, blue) and Figure \ref{fig:uves_spectral_atlas_red} (CD\#4, red) --- we achieve a median SNR of $\sim 1000$ in the blue arm, marginally above the expected $\sqrt{N}$ scaling owing to our use of sigma-clipping, but likely affected by line variability.
	
	\subsection{Doppler tomography}
	
\begin{figure*}
		\centering
		\includegraphics[width=\textwidth]{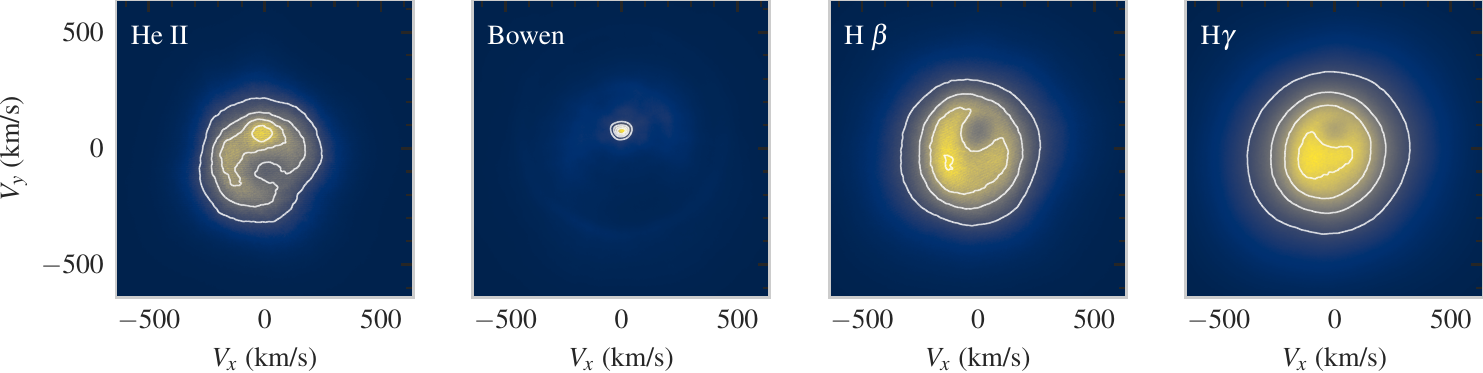}
		\caption{Doppler tomograms for prominent spectral features of interest in the UVES blue arm. Contours of constant intensity are overplotted to aid visualisation. Note that the Bowen image is dominated by the strong donor star features, with only tenuous hints of the sparse disc visible.}
		\label{fig:all_doppler_maps}
	\end{figure*}

	We applied Doppler tomography \citep{Marsh1988, Marsh2005} to our phase-resolved UVES spectra to probe the structure of the accretion disc surrounding Sco~X-1. Our spectra are pre-processed to remove continuum flux using the same procedure as Section~\ref{sec:lineatlas} and normalised in flux over the line region of interest to mitigate the impact of flux variability on the final Doppler map.
	We use the Python bindings of the \texttt{DOPPLER}\footnote{\url{https://github.com/trmrsh/trm-doppler}} code to produce the Doppler maps. The period and systemic velocity are fixed according to the median of our ephemeris posteriors (see Table \ref{tab:the_ephemeris}), and we compute the Doppler map over a square grid \SI{500}{\kilo\metre / \second} in side length, with a per-pixel resolution of \SI{1.5}{\kilo \metre / \second} in the x-y disc plane velocities $v_x$ and $v_y$. We neglect any velocities in the $z$ plane for computational reasons, but note that our spectra are unlikely to be constraining along this axis (see \citealt[]{Marsh2022}). Doppler maps are constructed for the H$\beta$, H$\gamma$, He~II and Bowen lines individually. These are shown in Figure \ref{fig:all_doppler_maps}, alongside the target $\chi^2$ value used for the maximum-entropy optimisation procedure. Despite the marked resolution improvement afforded by VLT/UVES over previous datasets, the output Doppler tomograms are of similar quality to the previous maps presented based solely on WHT data in \citet{Wang2018}. We speculate that this likely arises to long-term seasonal variations in the structure of the accretion disc around Sco~X-1, and explore this possibility further in the following section. 
	
	\subsection{Secular variability in the He II disc?} \label{sec:seasonaltomograms}
	
\begin{figure*}
		\centering
		\includegraphics[width=\linewidth]{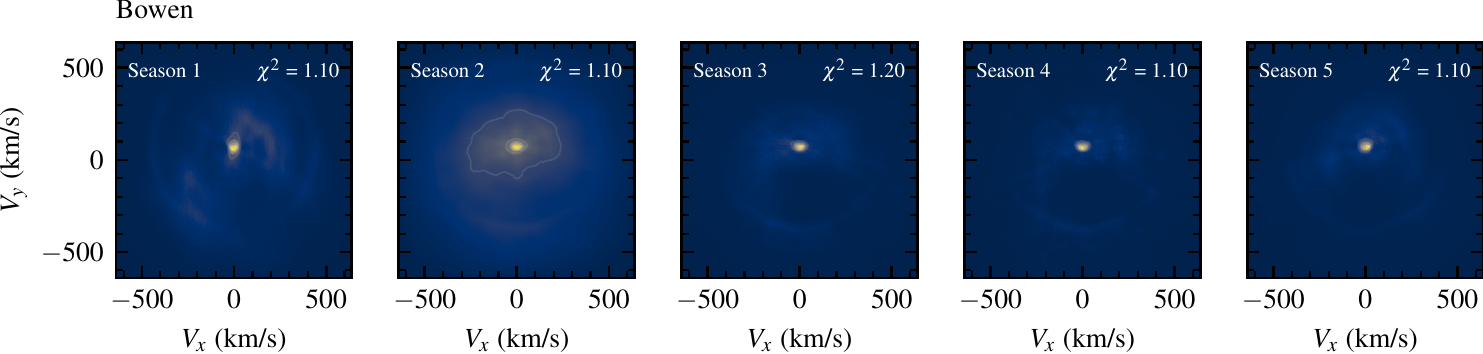}
		\includegraphics[width=\linewidth]{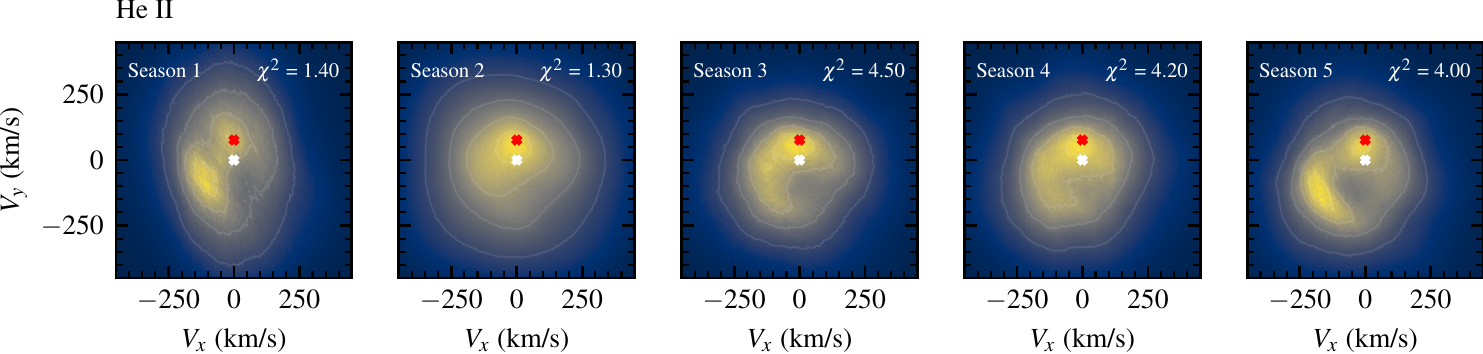}
		\caption{
			Doppler tomograms for each observing `season', using the method defined in Section~\ref{sec:seasonaltomograms}. We compute season-averaged tomograms for the Bowen lines (top panel) and the He~II lines(bottom panel), to search for long-term secular variations in the disc structure. The Bowen Doppler maps are largely dominated by the strong donor star signature, but some structure is visible. The red and white markers indicate the positions of the donor star, and the system centre-of-mass respectively.
		}
		\label{fig:seasonal_doppler_maps}
	\end{figure*}

	To explore the possibility of seasonal or secular variability in the disc structure of Sco~X-1, we compute He~II Doppler maps on subsets of the full UVES dataset presented in this paper. We manually group contiguous subsets of our data into five `seasons' -- selected roughly corresponding to the ESO observing semesters. We then compute individual Doppler maps for each season, to avoid averaging over any long-term variability. The target $\chi^2$ is adjusted per-season to avoid over-resolving disc features.
	The resulting Doppler maps are shown in Figure \ref{fig:seasonal_doppler_maps}.
	
	We focus our attention on the He~II Doppler maps, as these show a clearly-resolved disc structure. Some tenuous variability is present in the Bowen disc also, although this is hard to quantify or comment on phenomenologically, given that the donor star feature (at $V_x = 0$, $V_y= 76.4$ km/s) dominates the reconstruction.
	In the He~II maps, we see a strong asymmetric emission component present in the accretion disc. Across multiple observing seasons, we see evidence for evolution in intensity of this `rim'-like structure --- with minimal intensity in Season 2 and maximal intensity in Season 5. We interpret this feature as an extended region of gaseous material suspended above the disc and corotating with it --- potentially a signature of stream--disc overflow occurring in the system (see, e.g., \citealt{Kunze2001}). Given the high accretion rate of Sco~X-1, this is not surprising, as the accretion process is stochastic and variable. As our Doppler maps are averaged over many orbital periods, this is likely to be a longer-term secular variation rather than an effect correlated with the disc state. This may potentially coincide with periods of enhanced accretion onto the NS, as implied by short- ($\sim$ weeks--months) timescale fluctuations in the X-ray luminosity seen in MAXI light curves \citep{Hynes2016}. These disc states are well-known (e.g., \citealt[]{Scaringi2015}), but we do not resolve these with our sparsely-sampled VLT/UVES dataset.
	
	We encourage additional observations and characterisation of these secular variations in the disc structure, preferentially via intensive spectroscopic observations over blocks of a few nights, every few months. This ensures each block has adequate phase coverage, and minimises the blurring of structure caused by a more coarse observing program such as our VLT/UVES data. Simultaneous X-ray coverage is crucial to correlate structural changes with potential modulation of accretion rate, and to understand the disc state. As the prototypical LMXB, targeted studies of this phenomenon in Sco~X-1 have the potential to provide insight into the longer-term dynamics of accretion across the domain of X-ray binaries and other high accretion rate compact binaries.

	\subsection{Behaviour of the Balmer lines}
	Complex absorption features are present in the Balmer lines --- as reported by \citealt[]{Scaringi2015} --- reminiscent of P Cygni-style outflow profiles with red/blue absorption wings superimposed on a broad central emission feature. These occur antiphase with the overall optical brightness of the system (and with the Balmer line intensity itself), showing maximum absorption when the Balmer lines are strongest and no absorption when the Balmer lines are barely visible. Both red and blue components are visible, showing considerable variations in absorption depth. Given the strong time variability demonstrated in previous sections, it is challenging to probe this with our VLT/UVES dataset. Nevertheless, we present some broad overview properties below in the hope of encouraging further characterisation. Such profiles have also been noted previously in NIR spectra of the system on the Br$\gamma$ line, e.g. see \citet{Bandyopadhyay1999}.
	
\begin{figure}
		\centering
		\includegraphics[width=\linewidth]{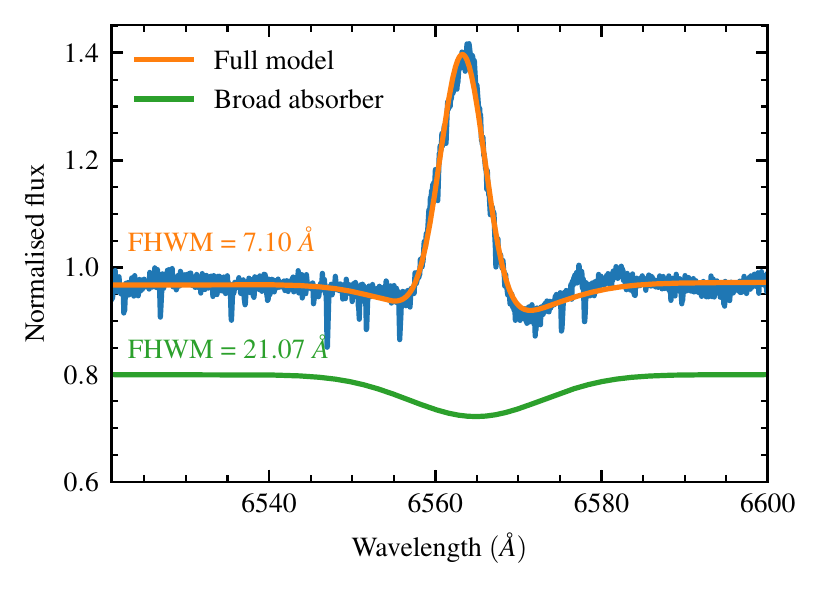}
		\caption{Spectrum of Sco~X-1~showing strong absorption wings on the H$\alpha$ line. Overplotted is a two-Gaussian model that reproduces the line shape shown, with one positive component representing the broad H$\alpha$ line, and one constrained-negative component reproducing the broad absorption wings present.}
		\label{fig:ha_absorption}
	\end{figure}

	We focus on the H$\alpha$ line as this shows the strongest absorption (see Figure \ref{fig:ha_absorption})and thus provides the most robust velocity measurements, although this phenomenon is also visible in H$_\beta,\gamma$. We apply a local continuum normalisation to bring all spectra onto a common baseline, then fit a Chebyshev polynomial to the line wing to provide a smooth approximation to the spectrum on this specific region. We find the minimum value of this polynomial and use bootstrap iterations to measure the uncertainty on our minimum wavelength value. To reject spectra where no absorption is present and thus ensure our velocity measurements are robust, we only consider spectra from this point on where there is $\geq$ 3\% absorption in this red wing.
	
	Under the assumption that this is an outflow, the `apparent' velocity varies between \SIrange{100}{1000}{\kilo\meter / \second}, with variable absorption depth as remarked upon above, and appears to show some correlation with orbital phase. However, such a phase-dependent morphology could equally be replicated with the addition of a broad absorption component superimposed on the emission component (see Figure~\ref{fig:ha_absorption}). To illustrate this, we fit a toy model to one of our spectra showing high absorption, composed of a narrow emission line centred on H$\alpha$, and a broad (at least 2x wider than the emission component) absorption line with a free mean value.  We note that the true structure of the Balmer lines is markedly more complex than this, and this simplified model is intended to be illustrative, rather than fully reproducing all features of the data.
	
	Owing to degeneracies between the strength of the $H\alpha$ line and this component it is not possible to constrain the amplitude of such absorption, but one solution suggests a width of around $8 \times$ that of $H\alpha$. This could be created by optically-thick absorption of the inner disk surrounding the NS in Sco~X-1, potentially probing some NS-driven outflow that may act as a crude tracer of the true NS radial velocity amplitude. This component appears to have some systemic velocity offset compared to the Balmer line in emission. It relies largely on serendipity to observe the disc in the right state to make these measurements, and would benefit strongly from simultaneous X-ray observations, to this end. Disentangling the origin is not possible with the sparse sampling of the VLT/UVES data, and we therefore advocate for additional intensive monitoring to investigate the nature of this absorption.
	
	\subsection{Constraints on other system parameters?}
	Despite the marked improvement in timing uncertainties offered by our reanalysis and extensive dataset, it is difficult to further constrain the system's orbital parameters, which are important in CW searches.
	The uncertainties on the other system parameters of interest are dominated by the poor constraints on the mass ratio $q$ and the unknown neutron star radial velocity $K_1$, that are difficult to further improve on.  The additional VLT/UVES data provide no stronger constraint on $K_1$ owing to the variability we discuss in Section~\ref{sec:seasonaltomograms} blurring the Doppler tomograms. Similarly, we cannot tighten the lower bound on $q$ as the lines are already fully resolved by our spectra.
	We recreated the Monte-Carlo analysis of \citet{Wang2018}, casting the problem instead as a Bayesian forward-modelling approach (using the HMC methods discussed in Section~\ref{sec: bayesian modelling}), but this failed to provide any stronger constraints, owing to the above considerations.
	
	Further observational work is required to better constrain these orbital parameters, yet the complexity of Sco~X-1 makes this challenging --- detection of donor star features in the near-infrared is hampered by the strong accretion flux \citep{MataSanchez2015}, and light-curve modelling is made more difficult by the distinct optical high and low states in the system. Only by jointly considering all available constraints, and pushing the capabilities of current-generation instruments, can we begin to more robustly constrain the orbital parameters of Sco-X~1 and further reduce the parameter space for CW searches going forwards.

	\section{Conclusions}
	\label{sec: conclusions}
	As the prototypical LMXB, Sco~X-1 continues to remain central to current searches for continuous GW sources. Tensions between the predicted GW emission and the GW strain limits obtained from previous observing runs are beginning to emerge, as specific sub-bands now reach below the torque--balance between accretion spin-up and GW emission spin-down. However, this is largely contingent on the inclination constraint obtained from \citet{Fomalont2001}, and further data is required to reach constraining upper limits whilst simultaneously marginalising over the unknown inclination; it may be the case, e.g., that the radio lobes do not trace the orbital inclination.
	
	Although any GW emission thus far has eluded detection, the upcoming LIGO/Virgo/KAGRA O4 observing run promises to further improve instrument sensitivities. Furthermore, searches with increased sensitivity (e.g., \citealt[]{2022arXiv220709326M}) may yield more stringent constraints over previous results. With upcoming third-generation GW detectors such as the Einstein Telescope \citep{Maggiore2020} and Cosmic Explorer \citep{2019BAAS...51g..35R} --- bringing orders of magnitude increase in strain sensitivity and delivering high SNR GW detections that will unveil populations of compact objects currently out of reach for the current ground-based detectors (e.g., \citealt{Cieslar2021}) --- likely including Sco~X-1, alongside a wealth of other science outcomes \citep{2021arXiv211106990K}.
	
	From a compact object perspective, further studies of Sco~X-1~focused on the transient structure of the disc and the nature of the companion star is crucial in underpinning our understanding of more distant LMXBs, and may yield even stronger constraints that may synergistically further constrain the parameter space for GW searches. It is clear many processes remain poorly understood.
	
	Leveraging the improved search sensitivity afforded by the enhanced detectors and computational methods discussed above is contingent on a precise  --- and, critically, up-to-date --- ephemeris for the donor star in Sco~X-1. The ephemeris presented in this work will enable precision searches for CW emission from Sco~X-1, which will further constrain any emission down to the torque--balance limit across the entire band of search frequencies --- both in the upcoming LIGO-Virgo-Kagra O4 observing run and beyond.
	Sparsely-sampled, high-resolution observations of Sco~X-1~over the coming years can efficiently keep this ephemeris current for the foreseeable future, underpinned and facilitated by the extensive and corrected VLT/UVES constraints that we present here.

	\section*{Data Availability}
	All reduced data (excluding those under a proprietary period) are publicly available via the ESO Archive and ING Archive respectively.
	All associated data and reduction codes will be made publicly available after the release of this manuscript at \url{https://github.com/tkillestein/pegs_plus}.
	We intend to regularly publish updates to this ephemeris in light of new data, with the version of record hosted at Zenodo. (DOI: 10.5281/zenodo.7635465)
	This paper makes use of the following software packages:
	\texttt{numpy} \citep{numpy},
	\texttt{scipy} \citep{scipy},
	\texttt{astropy} \citep{astropy1, astropy2},
	\texttt{matplotlib} \citep{matplotlib},
	\texttt{jax} \citep{jax2018github},
	\texttt{numpyro} \citep{phan2019composable, bingham2019pyro},
	\texttt{corner} \citep{corner},
	\texttt{pandas} \citep{pandas},
	\texttt{cividis} colour map \citep{cividis}

	\section*{Acknowledgements}
	
	We thank the anonymous referee for their careful review of our manuscript.
	Based on observations collected at the European Organisation for Astronomical Research in the Southern Hemisphere under ESO programmes 077.D-0384(A), 087.D-0278(A), 089.D-0272(A), 098.D-0688(A), 599.D-0353(A), and 599.D-0353(B). 
	and on observations made with the William Herschel Telescope operated on the island of La Palma by the Isaac Newton Group of Telescopes in the Spanish Observatorio del Roque de los Muchachos of the Instituto de Astrofísica de Canarias under programmes W1999A/CAT42 and W/2011A/P23 respectively.
	This paper makes use of data obtained from the Isaac Newton Group Archive which is maintained as part of the CASU Astronomical Data Centre at the Institute of Astronomy, Cambridge.
	
	T.L.K. gratefully acknowledges support fom the UK Science and Technology Facilities Council (STFC, grant number ST/T506503/1).
	M.M. is supported by European Union's H2020 ERC Starting Grant No. 945155--GWmining and Cariplo Foundation Grant No. 2021-0555.
	D.S. acknowledges support from the UK Science and Technology Facilities Council (STFC, grant numbers ST/T007184/1, ST/T003103/1 and ST/T000406/1).
	J.C. acknowledges support by the Spanish Ministry of Science under grant PID2020-120323GB-100.
	J.T.W. acknowledges support from NSF grants PHY-1806824 and PHY-2110460.
	
	\bibliographystyle{mnras}
	\bibliography{draft_dedupe}
	\bsp
	\label{lastpage}
\end{document}